\documentclass[aps,prl,10pt,twocolumn,showpacs,floatfix,superscriptaddress]{revtex4-1}
	\usepackage{graphicx}
	\usepackage{graphics}
	\usepackage{amsmath}
	\usepackage[usenames]{color}
	\usepackage{verbatim}
	\usepackage{braket}	
	\usepackage{cases}
\usepackage[breaklinks=true,colorlinks=true,linkcolor=blue,urlcolor=blue,citecolor=blue]{hyperref}
	
	\newcommand{\be}{\begin{equation}}
	\newcommand{\ee}{\end{equation}}
	\newcommand{\bea}{\begin{eqnarray}}
	\newcommand{\eea}{\end{eqnarray}}
	
\begin{document}
\title{Exact calculation of the probabilities of rare events in cluster-cluster aggregation}

\author{R. Rajesh}
\email{rrajesh@imsc.res.in}
\affiliation{The Institute of Mathematical Sciences, C.I.T. Campus, Taramani, Chennai 600113, India}
\affiliation{Homi Bhabha National Institute, Training School Complex, Anushakti Nagar, Mumbai 400094, India}

\author{V. Subashri}
\email{subashriv@imsc.res.in}
\affiliation{The Institute of Mathematical Sciences, C.I.T. Campus, Taramani, Chennai 600113, India}
\affiliation{Homi Bhabha National Institute, Training School Complex, Anushakti Nagar, Mumbai 400094, India}

\author{Oleg Zaboronski}
\email{olegz@maths.warwick.ac.uk}
\affiliation {Mathematics Institute, University of Warwick, Gibbet Hill Road, Coventry CV4 7AL, UK}
\date{\today}
		
\begin{abstract}
We develop an action formalism to calculate probabilities of rare events in cluster-cluster aggregation for arbitrary collision kernels and establish a pathwise large deviation principle with total mass being the rate. As an application, the rate function for the number of surviving particles as well as the optimal evolution trajectory are calculated exactly for the constant, sum and product kernels. For the product kernel, we argue that the second derivative of the rate function has a discontinuity. The theoretical results agree with simulations tailored to the calculation of rare events.
\end{abstract}

\maketitle

The study of cluster-cluster aggregation  (CCA), a nonequilibrium, irreversible phenomenon where particles, or clusters coalesce on contact to form larger clusters, has a long history dating back to Smoluchowski in 1917~\cite{smoluchowski1917mathematical}. It has been studied extensively because of its occurrence in diverse physical phenomena such as blood coagulation~\cite{guria2009mathematical}, cloud formation~\cite{falkovich2002acceleration,pruppacher2012microphysics}, aerosol dynamics~\cite{williams1988unified}, coagulation of dust and gas particles forming Saturn's rings~\cite{brilliantov2015size}, aggregation of particulate matter in oceans~\cite{burd2009particle}, protein aggregation~\cite{benjwal2006monitoring,wang2015following}, charged biopolymers~\cite{tom2016aggregation,tom2017aggregation}, ductile fracture~\cite{pineau2016failure}, etc. CCA also finds applications in applied fields such as river networks~\cite{tarboton1988fractal}, mobile networks~\cite{heimlicher2010globs}, population genetics~\cite{berestycki2009recent} and explosive percolation~\cite{achlioptas2009explosive,d2019explosive},etc.
	
CCA has been analyzed using different approaches. The most common approach is to solve the deterministic mean-field Smoluchowski equation that describes the change in the number of clusters of a given mass due to coagulation events (see Refs.~\cite{leyvraz2003scaling,aldous1999deterministic, krapivsky2010kinetic,wattis2006introduction} for reviews). The Smoluchowski equation for the mean mass distribution is exactly solvable when the rate of collision is independent of the masses (constant kernel), is the sum of the masses (sum kernel), and product of the masses (product kernel). For the product kernel, a sol-gel transition is observed wherein the total mass is not conserved beyond a gelling time. For the sum kernel, the gelling occurs at infinite time~\citep{krapivsky2010kinetic}. In lower dimensions, spatial density fluctuations become important and have been studied using both analytical and numerical techniques~\cite{spouge1988exact,kang1984fluctuation,krishnamurthy2002kang,krishnamurthy2003persistence}. These approaches are, however, restricted to studying the mean or typical mass distribution and the low order moments of the mean mass distributions and do not give information about either the probabilities of rare or atypical events or the trajectories that lead to atypical events.  In this paper, we present an exact calculation of these probabilities.

The tails of a probability distribution describe events, which while rare, are important to study because their impact may be significant. Examples of impactful rare events include heat waves~\cite{ragone2021rare,ragone2018computation}, earthquakes~\cite{ben2020localization}, extreme events in climate and ecosystems, such as the loss of sea ice in the Arctic region ~\cite{overland2021rare}, etc. In particular, examples of rare events in CCA include neurological disorders such as Alzheimers' disease~(\cite{iannucci2020thrombin}), mad cow disease~\cite{nowak1998prion}, the clustering of raindrops leading to rapid onset of rainfall~\cite{wilkinson2016large}, etc. The probabilities of rare events is captured by the large deviation function (LDF) or the rate function, and falls into the general framework of large deviation theory. The LDF can be interpreted as a nonequilibrium generalization of entropy or free energy.
		
Consider a collection of massive particles which evolves in time through  binary mass-conserving aggregation (also known as Marcus-Lushnikov model~\cite{marcus1968stochastic,
lushnikov2006gelation,
lushnikov1973evolution,
lushnikov1978coagulation}):
	\begin{equation}
	A_i + A_j  \xrightarrow{\lambda K(i,j)} A_{i+j},
	\end{equation}
where $A_k$ denotes a particle of mass $k$ and $\lambda K(i,j)$ is the rate at which two particles of masses $i$ and $j$ aggregate.  We note that all the spatial information has been encoded into the collision kernel, $K(i,j)$. Let $N(t)$ denote the number of particles at time $t$.  Initially, there are $N(0)=M$ particles of equal mass (set equal to $1$). A quantity of interest is the probability density function $P(M,N,t_f)$, of having exactly $N$ particles remain at time $t_f$. Additionally,  we ask what the most probable trajectory is  for a given  $M, N, t_f$.

In this paper, we study the LDF for CCA using the Doi-Peliti-Zeldovich (DPZ) method~\cite{doi1976second,doi1976stochastic,peliti1985path,ovchinnikov1978role,tauber2014,cardy2006reaction},  a path integral method. The LDF  is calculated exactly for the constant, sum and product kernels. For the product kernel, we argue that the LDF is singular with a discontinuity in the second derivative, indicating the sol-gel transition. Gelation transition has been studied using large deviation theory in the probability literature (see~\cite{andreis2019large} and references within). These results are based on special factorisation properties of the law of the mass distribution. In contrast, we derive the pathwise large deviation principle which, at least formally, is valid for arbitrary kernels. This more general point of view allows us to determine the optimal evolution trajectories as solutions to the Euler-Lagrange equations for effective action functional.

We first express $P(M,N,t_f)$ in terms of an effective action~\cite{connaughton2006cluster,connaughton2005breakdown}. Let $\widetilde{P}(\vec{N},t)$ denote the probability of a system being in a configuration $\vec{N}$ at time $t$, where $\vec{N}(t)=\{N_1(t), N_2(t),  \ldots N_{M}(t)\}^{\mathsf{T}}$,  and $N_i(t)$ is the number of particles of mass $i$ at time $t$. Then,
	\begin{equation}
	P(M,N,t_f)=\sum_{\vec{N}}\widetilde{P}(\vec{N},t_f)\delta\left(\sum_{i=1}^{M}N_i(t_f)-N\right).
\label{P}
	\end{equation} 
The time evolution of  $\widetilde{P}(\vec{N},t)$ is described by the master equation:
\begin{align}
&\frac{d \widetilde{P}(\vec{N})}{dt} \!=\!\!\sum_{i,j}\! \frac{\lambda K(i,j)}{2}\Big[(N_{i}\!+\!1\!+\!\delta_{i,j})(N_{j}\!+\!1)\nonumber\\&\widetilde{P}(\vec{N}\!+\!\mathcal{I}_{i}\!+\!\mathcal{I}_{j}\!-\!\mathcal{I}_{i+j})\!-\!N_{i}(N_{j}\!-\!\delta_{i,j})\widetilde{P}(\vec{N})\Big],
\label{master}
\end{align}
where $\mathcal{I}_k$ is the $M$-dimensional column vector whose $j$-th component equals $\delta_{jk}$. The first term in the right hand side of Eq.~(\ref{master}) enumerates all possible collisions that lead to  $\vec{N}$ while the second term enumerates all possible collisions that  lead to the system exiting $\vec{N}$.

The DPZ formalism allows one to rewrite the master equation in the form of a Schroedinger equation in imaginary time. The corresponding effective Hamiltonian is a polynomial in annihilation and creation operators $a_m,a_m^{\dagger}$ of particles of mass $m\geq 1$. These satisfy the canonical commutation relations $[a_m,a_n^{\dagger}]=\delta_{mn},[a_m,a_n]=[a_m^{\dagger},a_n^{\dagger}]=0$. Using the Trotter formula and  complete sets of coherent states, a solution to the master equation can be constructed in the form of a path integral. In particular, the probability $P(M,N,t_f)$, after  substituting $\phi=N/M$ and  $\tau=M\lambda t_f$, can be written as  (see \cite{subashrisupp} for the derivation)
\begin{equation}
	P(M,N,t_f)=\!\!\sideset{}{'}\sum_{k_i=1}^{k^*}\!\!\!\!\int \!\!\mathcal{D}\tilde{z}_i \mathcal{D}z_i \!\!\prod_{n=1}^N z_{k_n}(\tau_f)e^{-M S(\phi, \tau_f;\{z_i,\tilde{z}_i\})},
\label{eq:actionfunctional}
	\end{equation}
where $k^*=M-N+1$, and $\prime$ denotes the constraint $\sum_{i} k_i=M$. The action $S$ is given in terms of the effective Hamiltonian $H$ as
\bea
&&S=\!\!\!\int_{0}^{\tau_f}\! \!d\tau \!\!\!\left[ \sum_{m=1}^M \tilde{z}_m\dot{z}_m\!\!+\!\!H(\{z_i,\tilde{z}_i\})\right]\!\!-\ln \tilde{z}_1(0)+1, \label{eq:action}\\
&&H(\{z_i\},\{\tilde{z}_i\})=-\frac{1}{2}\sum_{i,j}K(i,j)(\tilde{z}_{i+j}-\tilde{z}_i\tilde{z}_j)z_i z_j.\label{eq:energy}
\eea
The action is invariant under the transformation $z_m\to c^m z_m$ and $\tilde{z}_m\to c^{-m}\tilde{z}_m$. Hence, we can set $\tilde{z}_1(0)=1$. 

In the limit $M \to \infty$, keeping $\phi$ and $\tau_f$ fixed, the functional integral in Eq.~(\ref{eq:actionfunctional}) is dominated by the minimum of $S$, and hence can be calculated using Laplace method. The corresponding Euler-Lagrange equations for 
${z}_m, \tilde{z}_m$, $m=1\dots M$, are
\begin{align}
	&\frac{d{z}_m}{d\tau}=\frac{1}{2}\sum_j K (m,j)z_j z_{m-j}-\sum_j K(m,j)\tilde{z}_jz_m z_j,\label{z}\\
&\frac{d\tilde{z}_m}{d\tau}=-\sum_j K(m,j)(\tilde{z}_{m+j}-\tilde{z}_m\tilde{z}_j)z_j,\label{tz} 
\end{align}
with the boundary conditions $\tilde{z}_m(\tau_f)z_m(\tau_f)=n_m(\tau_f)$ and $z_1(0)\tilde{z}_1(0)=1$, where $M n_m(\tau)$ is the number of particles of mass $m$ at time $\tau$. This is consistent with the DPZ formalism, where $n_m= z_m \tilde{z}_m$. The time evolution of $n=\sum_i z_i\tilde{z}_i$, the fraction of particles, is then given by
	\begin{equation}
	\frac{dn}{d\tau}=\frac{-\sum_{i,j} K(i,j) n_i n_j}{2} + E,\!
	\! ~n(0) = 1,n(\tau_f)=\phi.
	\label{inst}
	\end{equation}

For $\tilde{z}_i$, $z_i$ satisfying the Euler Lagrange equations, it can be shown that $H$ reduces to $2 H= \sum_i z_i \frac{d\tilde{z}_i}{d\tau}$. Also, $H$ is a constant of motion  (see~\cite{subashrisupp} for the proof), and we denote its value by $E$.  We note that Eq.~(\ref{tz}) is satisfied by $\tilde{z}_i(\tau)=1$, in which case, $E=0$ [see Eq.~(\ref{eq:energy})]. For this special case, Eq.~(\ref{z}) for $z_m$, and hence $n_m$,  is identical to the Smoluchowski equation for the mean mass distribution, and thus will correspond to the typical solution for a given time. We now discuss the general case, $E\neq 0$, corresponding to atypical solutions. Evaluating the integral Eq.~(\ref{eq:action}), we then obtain
\be
\label{eq:modac1}
P(M,\!N,\tau_f)\!\sim\! \!\sideset{}{'}\sum_{k_i=1}^{k^{*}}\!\prod_{n=1}^N \!z_{k_n}(\tau_f)e^{-M\left[\phi\ln\phi-E\tau_f+1\right]}
\ee
Equations~(\ref{z}), (\ref{tz}) and (\ref{eq:modac1}) describe the calculation of the LDF for an arbitrary kernel. Since $N=\phi M$, it is clear that in the limit $M \to \infty$, keeping $\phi$ and $\tau_f$ fixed, we can define a LDF
\be
f(\phi, \tau) = \lim_{M\to \infty} \frac{-1}{M} \ln P(M,M \phi,\tau/M),
\label{eq:fdefinition}
\ee
thus establishing a large deviation principle for any collision kernel. 
We will now present an exact calculation of $f(\phi,\tau)$ for the constant, sum and product kernels.

{\em Constant kernel} [$K(i,j)=1$]:
The instanton equation, Eq.~(\ref{inst}), reduces to $dn/d\tau=-n^2/2+E$. Since $n(\tau)$ decreases with time, $E<n^2/2$.
The solution for $n(\tau)$ is (see \cite{subashrisupp} for more details)
\begin{equation}
n(\tau)=\begin{cases}
-\sqrt{-2E}\tan\frac{\sqrt{-2E}(\tau-\tau_0)}{2}, & E<0,\\
\frac{1}{1+\tau/2},			& E=0,\\
\sqrt{2E}\coth \frac{\sqrt{2E}(\tau-\tau_1)}{2}, & E>0,
\end{cases}\label{eq:n}
\end{equation}
where the constants $E,\tau_0,\tau_1$ are fixed by the boundary conditions in Eq.~(\ref{inst}). For determining the LDF, we also need to determine $z_m(\tau_f)$ and $\tilde{z}_1(0)$. Writing $z_m(\tau)$ in terms of its generating function, $Y(x,\tau)=\sum_m z_m(\tau)x^{m}-n(\tau)$, we obtain
\begin{equation}
\frac{\partial Y}{\partial \tau}=\frac{Y^2}{2}-E,~~Y(x,0)=y_1(0)x.\label{eq:Y}
\end{equation}
Solving for $Y$ and hence $z_m(\tau)$, we obtain
\begin{equation}
z_m(\tau)=\begin{cases}
\frac{-2Ez_1(0)^m \sec^2\sqrt{-E/2}\tau\left[\tan\sqrt{-E/2}\tau\right]^{m-1}}{\left[\sqrt{-2E}+\tan\sqrt{-E/2}\tau\right]^{m+1}}, &\!\!\!\! E<0,\\
\frac{4\tau^{m-1}(z_1(0))^m}{(2+\tau)^{m+1}},&\!\!\!\! E=0,\\
\frac{2E z_1(0)^m \sinh^{m-1}\sqrt{E/2}\tau}{\left[\sinh\sqrt{E/2}\tau+\sqrt{2E}\cosh \sqrt{E/2}\tau\right]^{m+1}}, &\!\!\!\! E>0.
\end{cases}\label{eq:z}
\end{equation}

The combinatorial prefactor in Eq.~(\ref{eq:modac1}) is easily done to be $\sideset{}{'}\sum_{k_i=1}^{k^*} 1={M-1\choose N-1}.$ Using Stirling's approximation and substituting for $z_m(\tau)$ in Eq.~(\ref{eq:modac1}), the LDF is 
\begin{align}
&f(\phi,\tau)=\\&
\begin{cases}
\phi\ln\frac{\phi^2}{-2E+\phi^2}+\ln(1-2E)-E\tau, \!\!&\!\!\! E<0,\\
0,\!\!& \!\!\!E=0,\\
-E\tau-2\phi\ln\frac{2E}{\phi}-(1-\phi)\ln\frac{\sinh\tau\sqrt{E/2}}{1-\phi}+\\(1+\phi)\ln (\sqrt{2E}\cosh\tau\sqrt{E/2}
+\sinh \tau\sqrt{E/2}), \!\!&\!\!\! E>0,\nonumber
\end{cases}\label{eq:z}
\end{align}
where $E<0$, $E=0$, $E>0$ correspond to final times $\tau<\tau_{typ}$, $\tau=\tau_{typ}$ and $\tau>\tau_{typ}$ respectively, and $\tau_{typ}$ is the typical time for the fraction of particles to reach $\phi$.
\begin{figure}
	\centering
	\includegraphics[width=\linewidth]{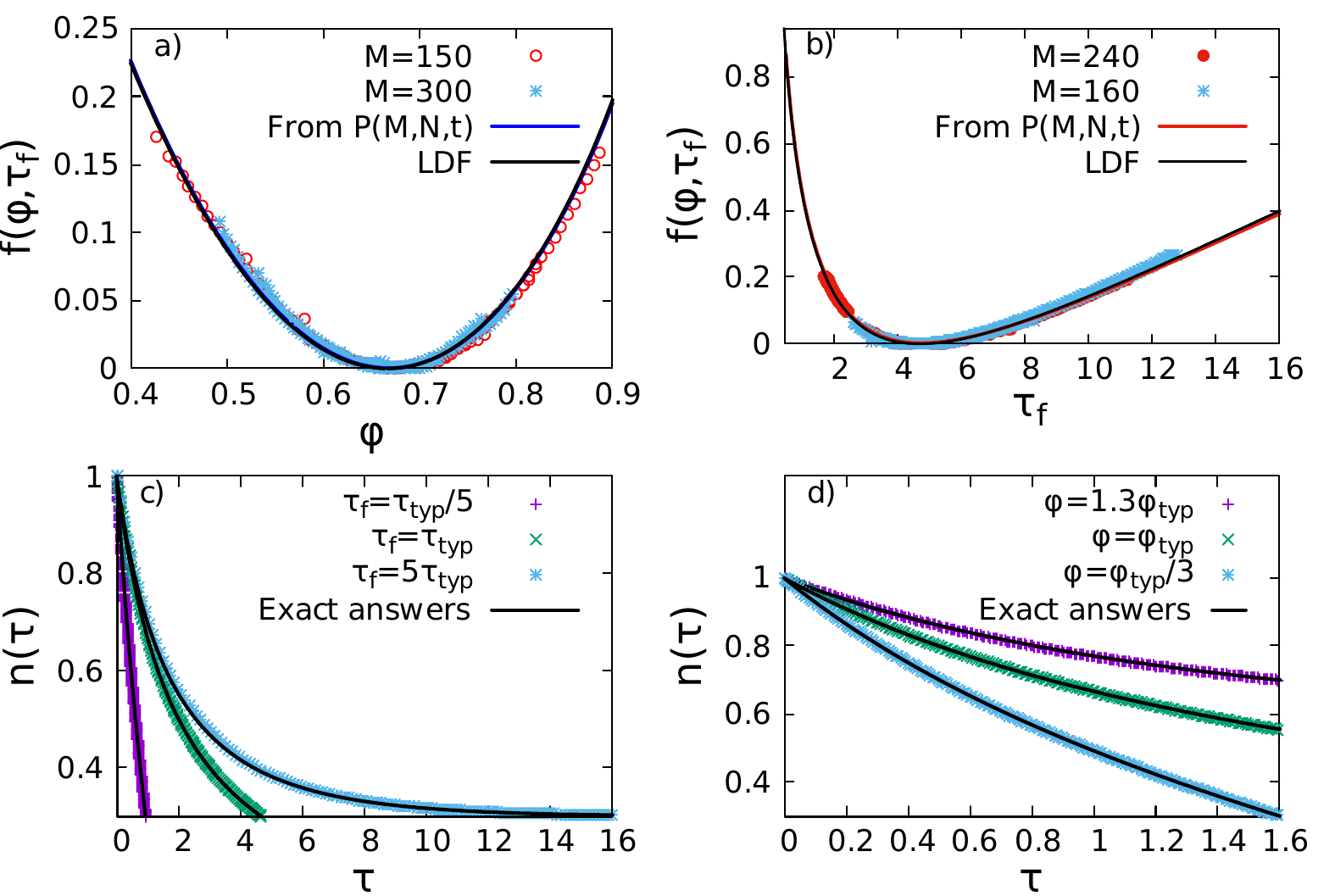}
	\caption{Constant kernel: Comparison of $f(\phi,\tau)$ with simulation data and exact expression for $P(M,N,t)$ for (a) varying $\phi$ for fixed $\tau_f=1$, (b) varying $\tau_f$ for $\phi=0.3$. The instanton trajectory in Eq.~(\ref{eq:n}) is compared with simulation data  for (c) $\phi=0.3$ and different $\tau_f$ and (d) $\tau_f=1.6$ and different $\phi$.
	}
	\label{fig:01}
\end{figure}

We demonstrate the correctness of the solution as well as the procedure by comparing  $f(\phi, \tau)$ with results from both Monte Carlo simulations and the exact expression for $P(M,N,t)$. The simulations are based on a biased Monte Carlo scheme for fixed number of particles~\cite{dandekar2023monte} that accurately determines the probabilities of rare events and the instanton trajectory. We generalise the algorithm to allow for number of particles to fluctuate (see \cite{subashrisupp} for more details). For the constant kernel, the reaction rate does not explicitly on the mass distribution and hence it is possible to write $P(M,N,t)$ as a sum over exponentials~\cite{dandekar2023monte}. We note that it is difficult to extract the LDF from this expression, however, it can be evaluated numerically.  We find an excellent agreement of $f(\phi, \tau$) with the simulations and exact answer both for fixed $\tau$ and varying $\phi$ [see Fig.~\ref{fig:01}(a)],  and fixed $\phi$ and varying $\tau$ [see Fig.~\ref{fig:01}(b)].  The analytical results for the instanton solution [see  Eq.~(\ref{eq:n})] are also in excellent agreement with the numerical results  for short, typical and long times [see Fig.~\ref{fig:01}(c), (d)].
\begin{figure}
	\includegraphics[width=\linewidth]{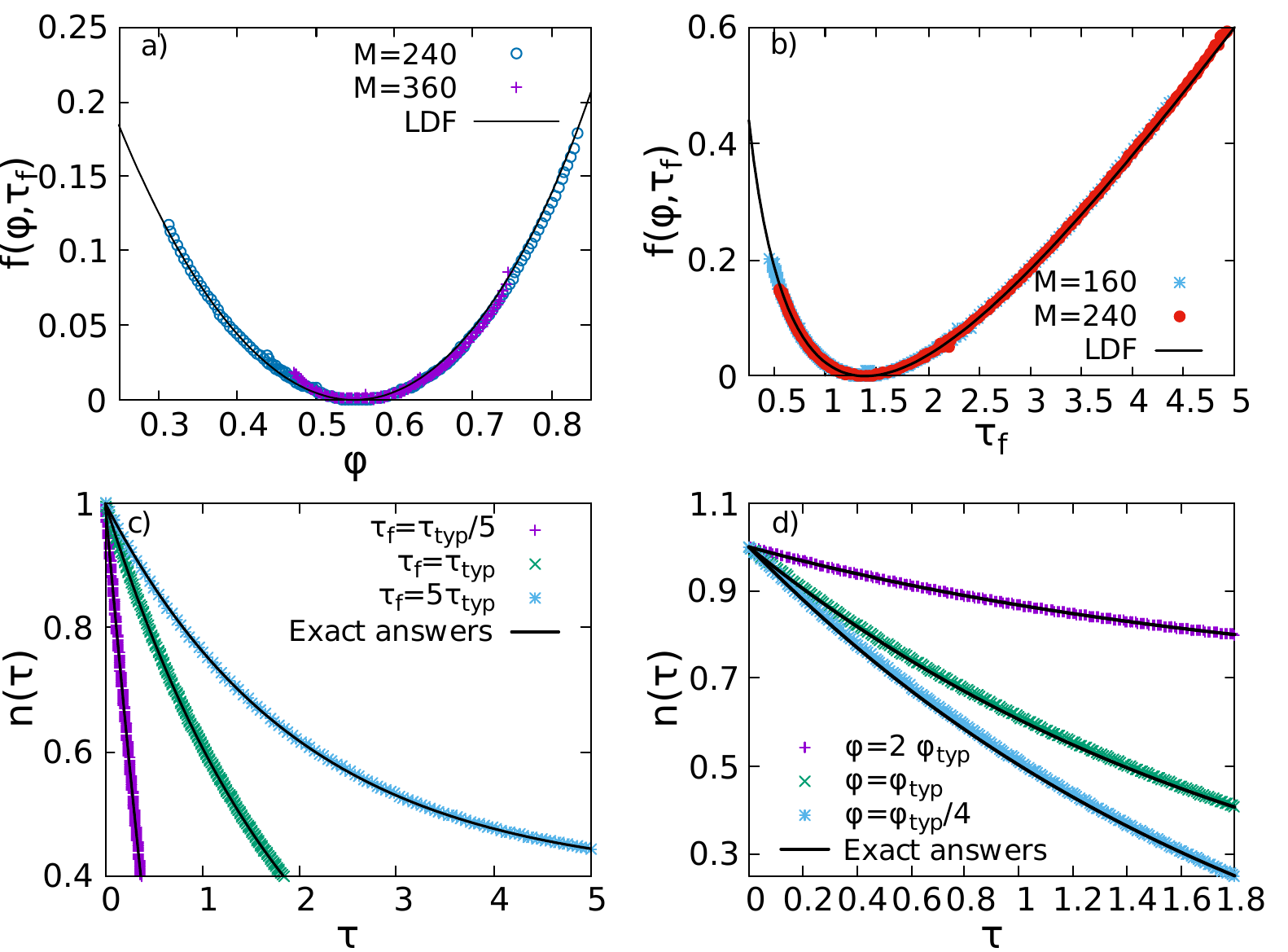}
	\caption{Sum kernel: Comparison of $f(\phi,\tau)$ with simulation data for (a) varying $\phi$ for fixed $\tau_f=1.2$, (b) varying $\tau_f$ for $\phi=0.5$. The instanton trajectory in Eq.~(\ref{eq:n-sum}) is compared with simulation data  for (c) $\phi=0.4$ and different $\tau_f$ and (d) $\tau_f=1.8$ and different $\phi$.}
	\label{fig:02}
\end{figure}

{\em Sum kernel} [$K(i,j)=(i+j)/2$]: The instanton equation, Eq.~(\ref{inst}),  reduces to $dn/d\tau=E-n/2$, with solution
\begin{equation}
n(\tau)=\frac{\phi-e^{-\tau_f/2}}{1-e^{-\tau_f/2}}-\bigg(\frac{\phi-1}{1-e^{-\tau_f/2}}\bigg)e^{-\frac{\tau}{2}}.
\label{eq:n-sum}
\end{equation}
The Euler-Lagrange equations  for $z_i$ [see Eq.~(\ref{z})] can now be solved (see \cite{subashrisupp}) to give
\begin{equation}
z_i(\tau)=\frac{i^{i-1}a_1^i}{i!}(1-e^{-\tau/2})^{i-1}e^{-\int d\tau^{\prime} \frac{in+1}{2}},\label{eq:sumz}
\end{equation}
where $a_1$ is a constant.
The combinatorial prefactor in Eq.~(\ref{eq:modac1}) is then 
\begin{align}
\!\!\!\sideset{}{'}\sum_{k_i=1}^{k^*} \prod_{n=1}^{N}\frac{k_n^{k_n-1}}{k_n!}\!\!=\!\!M\bigg[\!-\!\!(1-\phi)\ln \frac{2(1-\phi)}{e^2}+\phi\ln \phi\bigg].\label{eq:sumcomb}
\end{align}
Substituting $z_m(\tau)$ and $I$ in Eq.~(\ref{eq:modac1}), we obtain LDF  for sum kernel to be 
\begin{equation}
f(\phi,\tau)=-(1-\phi)\ln \frac{1-e^{-\frac{\tau}{2}}}{1-\phi}+\frac{\tau\phi}{2}+\phi\ln\phi.\label{eq:sum}
\end{equation}

We find an excellent agreement of $f(\phi, \tau$) with the simulations both for fixed $\tau$ and varying $\phi$ [see Fig.~\ref{fig:02}(a)],  and fixed $\phi$ and varying $\tau$ [see Fig.~\ref{fig:02}(b)].  The analytical results for the instanton solution [see  Eq.~(\ref{eq:n-sum})] are also in excellent agreement with the numerical results  for short, typical and long times [see Fig.~\ref{fig:02}(c), (d)].

{\em Product Kernel:}
For the product kernel  the Smoluchowski equation predicts that a gel that contains a finite fraction of the mass forms at gelling time $\tau_g=1$ and gelling density $\phi_g=0.5$. In the discussion following  Eq.~(\ref{inst}), we showed that $E=0$ corresponds to the solution to the Smoluchowski equation. However, this solution  cannot be correct  for $\tau\geq 1$ as mass is not conserved, violating the strict conservation of mass in the Marcus-Lushnikov model. We, therefore, modify the solution for product kernel as follows.
\begin{figure}
	\includegraphics[width=\linewidth]{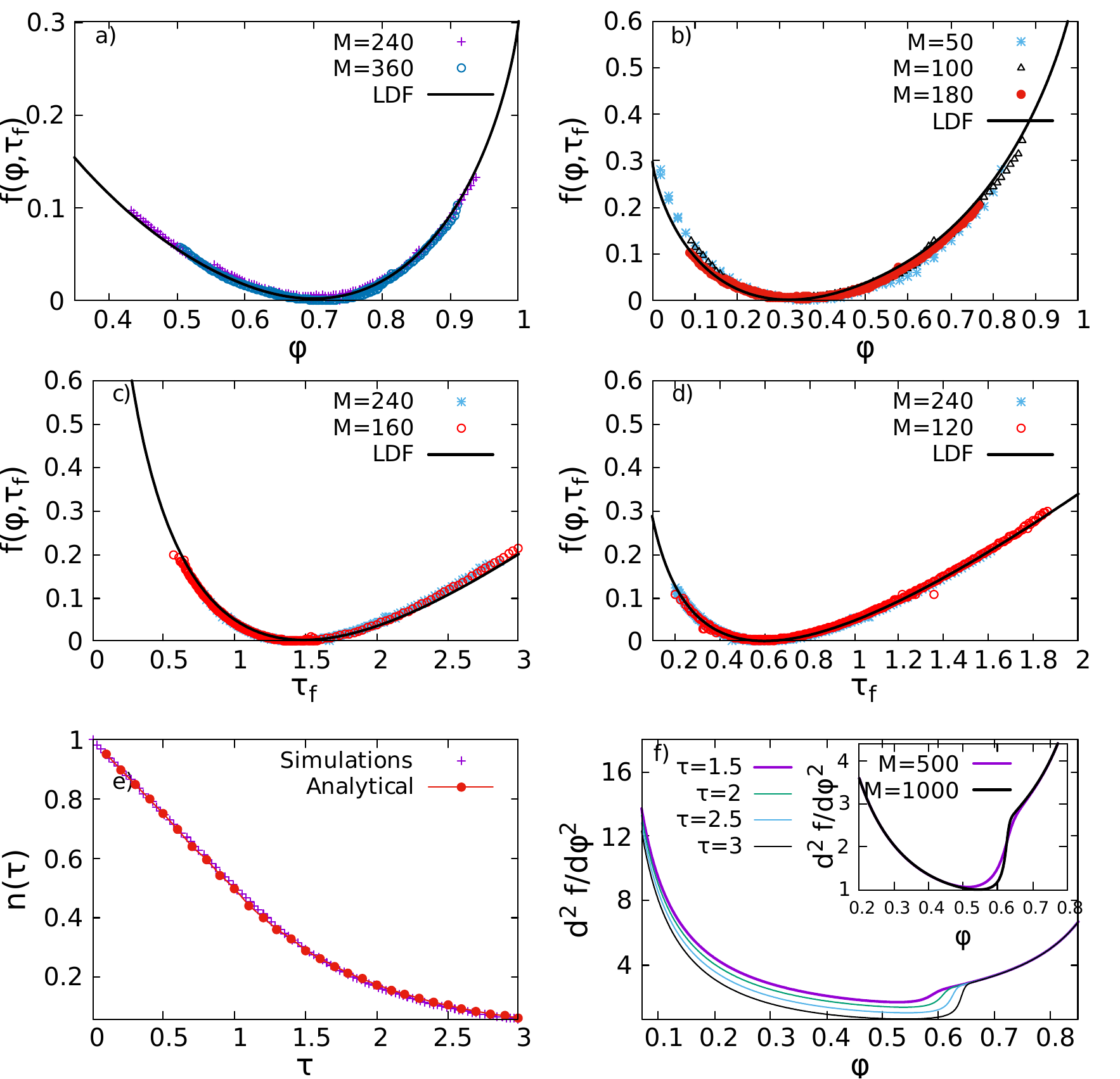}
	\caption{Product kernel: Comparison of $f(\phi,\tau)$ with simulation data for (a) fixed $\tau_f=0.6<\tau_g$, (b)  fixed $\tau_f=1.4>\tau_g$ (c) fixed $\phi=0.3<\phi_g$ and (d) fixed $\phi=0.7>\phi_g$. (e) The value of $\phi$ at minimum of $f(\phi, \tau)$ is compared with Monte Carlo simulations of the typical trajectory. (f) $d^2f/d\phi^2$ is discontinuous with $\phi$. inset: the discontinuity becomes sharper with increasing $M$. }
	\label{fig:03}
\end{figure}

We rewrite the unscaled Hamiltonian using  number operator $\hat{n}_i$ and total mass operator $\hat{M}$, breaking normal ordering. 
To restore normal ordering, we use the relation $\sum_i\hat{M}\ket{\psi(0)}=M\ket{\psi(0)}$, where $\ket{\psi(0)}=a_1^{\dagger M}\ket{\vec{0}}$, to rewrite $P(M,N,t_f)$ as
\begin{eqnarray}
\!\!P(M,N,t_f)\!\!=\!\!\bra{\vec{N}}\frac{(\sum_i a_i)^N}{N!} e^{-H^{\prime}(\{a^{\dagger}_i\},\{a_i\})t_f}\ket{\psi(0)},\\
H^{\prime}=-\frac{1}{2}\sum_i \sum_j ija^{\dagger}_{i+j}a_ia_j+ \sum_j\frac{(Mj-j^2)a^{\dagger}_ja_j}{2}.\label{eq:H'}
\end{eqnarray}
On introducing coherent states, we obtain the Euler-Lagrange equations to be
\begin{align}
\dot{z}_k&=\frac{1}{2}\sum_{l=1}^{k^{*}}l(k-l)z_lz_{k-l}-Mkz_k+\frac{k^2z_k}{2}\label{eq:pz},\\
\dot{\tilde{z}}_k&=-\sum_{l=1}^{k^{*}} kl \tilde{z}_{l+k}z_l+kM\tilde{z}_k-\frac{k^2\tilde{z}_k}{2}.\label{eq:pzbar}
\end{align}
We note that we could have followed the same procedure of introducing the operators $\hat{M}$ and $\hat{n}_i$ for the constant and sum kernels. For these kernels, the extra terms are always subleading in $M$ and thus, we obtain the same LDF. However, for the product kernel, the extra terms become important when a gel is present, and hence cannot be neglected.

Equation~(\ref{eq:pz}) can be solved exactly. Let $G(x,t)=\sum_{m=1}^M z_m(t)x^m$. Then, making the Cole-Hopf transformation $G(x,t)=\ln D(p(x,t),t)$, where $p(x,t)=xe^{-Mt}$, and solving the resulting partial differential equation using Knuth identity~\cite{knuth1998linear} (see \cite{subashrisupp} for details), we obtain
$z_m$ to be 
\begin{equation}
z_m(\tau)=\frac{(e^{\tau/M}-1)^{m-1}F_{m-1}(e^{\tau/M})M^{m-1}e^{-m \tau}}{m!},
\end{equation}
where $F_m(x)$ are the Mallows-Riordan polynomials~\cite{knuth1998linear, lushnikov2006gelation}. From Eqs.~(\ref{eq:pz}) and (\ref{eq:pzbar}),  we find that $\sum_i \dot{z}_i\tilde{z}_i=-E^{\prime}-M^2/2$, where $E'$ is the value of $H'$.  Substituting for $z_i$ in Eq.~(\ref{eq:modac1}), and computing the combinatorial prefactor,  we obtain the LDF for the product kernel:  
\begin{align}
f(\phi,\tau)&=\ln \frac{\phi^\phi e^{\tau/2+1-\phi}}{\tau^{1-\phi}}
+\min_x \{\ln x-\phi~h(x)\},\\
h(x)&=\sum_{k=1}^{k^{*}}\frac{x^k F_{k-1}(e^{{\tau}/{M}})}{k!}.
\end{align}

We find an excellent agreement of $f(\phi, \tau$) with the simulations for both pre-gelling and post-gelling regimes  [see Fig.~\ref{fig:03}(a)-(d)].  We also confirm that the minimum of the action corresponds to the typical solution [see Fig.~\ref{fig:03}(e)]. Finally, we find that the $\partial^2 f/\partial^2\phi$ has a discontinuity at a critical $\phi$ [see Fig.~\ref{fig:03}(f)]. The discontinuity becomes sharper with $M$ [see inset of Fig.~\ref{fig:03}(f)], suggesting the presence of a second order phase transition.

In summary, we developed a formalism to calculate the probabilities of rare events in cluster-cluster aggregation and demonstrated the existence of a large deviation principle for any collision kernel. The LDF is calculated exactly for the constant, sum, and product kernels. The known sol-gel transition for the product kernel is reflected as  a singular behaviour in the LDF. Our general method allows us to obtain the optimal evolution trajectory corresponding to any rare event.  These exact solutions will serve as a guideline for the numerical investigation of rare events in aggregation with collision kernels applicable to particular physical systems.

%

\end{document}


\title{
Exact calculation of the probabilities of rare events in cluster-cluster aggregation
}
\author{R. Rajesh}
\email{rrajesh@imsc.res.in}
\affiliation{The Institute of Mathematical Sciences, C.I.T. Campus, Taramani, Chennai 600113, India}
\affiliation{Homi Bhabha National Institute, Training School Complex, Anushakti Nagar, Mumbai 400094, India}

\author{V. Subashri}
\email{subashriv@imsc.res.in}
\affiliation{The Institute of Mathematical Sciences, C.I.T. Campus, Taramani, Chennai 600113, India}
\affiliation{Homi Bhabha National Institute, Training School Complex, Anushakti Nagar, Mumbai 400094, India}

\author{Oleg Zaboronski}
\email{olegz@maths.warwick.ac.uk}
\affiliation {Mathematics Institute, University of Warwick, Gibbet Hill Road, Coventry CV4 7AL, UK}

\date{\today}

\begin{center} \bf{Supporting Information}
\end{center}
\maketitle

\beginsupplement

		\section{Doi-Peliti-Zeldovich method}\label{1}
		\subsection{Derivation of effective action}
		We describe the methodology used to derive the probability of interest, $P(M,N,t)$ from the master equation [Eq.~(\ref{master}) of the main text] in terms of a large deviation function, for an arbitrary collision kernel $K(i,j)$. Using the Doi-Peliti-Zeldovich (DPZ) approach, we first express $P(\vec{N})$ as a path integral. 
		
The number of clusters of mass $i$, $N_i$, is denoted as the eigenvalue of a number operator $\widehat{N}_i$ acting on a state $\ket{\vec{N}}=\ket{N_1,N_2,\dots N_{M}}$, 
	\begin{equation}
	{\widehat{N}_i}\ket{\vec{N}}=N_i\ket{\vec{N}}. 
	\end{equation}
	The operator $\hat{N}_i$ is expressed in terms of annihilation and creation operators $a_i$ and $a_i^{\dagger}$, as
	\begin{equation}
	\hat{N}_i=a^{\dagger}_ia_i.
\end{equation}	
The annihilation and creation operators have the following properties,
	\begin{align}
	&a_i\ket{\vec{N}}=N_i\ket{\vec{N}-\mathcal{I}_i},\\
	&a^{\dagger}_i\ket{\vec{N}}=\ket{\vec{N}+\mathcal{I}_i},\\
	&a^{\dagger}_i{a}_i\ket{\vec{N}}=N_i\ket{\vec{N}}\label{seq:N},
	\end{align}
	where $\mathcal{I}_i$ denotes the change in $\ket{\vec{N}}$ through the increase or decrease of the number of clusters of mass $i$ by $1$. The commutations of the creation and annihilation operators have been mentioned in the main text. 
	A state $\ket{\psi(t)}$ is defined as a linear combination of $\ket{\vec{N}}$:
	\begin{equation}
	\ket{\psi(t)}=\sum_{\vec{N}}P(\vec{N},t)\ket{\vec{N}}.\label{master1}
	\end{equation}
	
 By multiplying both sides of the master equation by $\ket{\vec{N}}$ and summing over the configurations, we obtain a differential equation for $\ket{\psi(t)}$. Further, using Eq.~(\ref{seq:N}), we obtain the master equation in the form of a Schroedinger equation,
	\begin{equation}
	\frac{d\ket{\psi(t)}}{dt}=-{H}(\{a^{\dagger}_i,a_i\}),\ket{\psi(t)},\label{sch}
	\end{equation}
	where the corresponding Hamiltonian $\widehat{H}(\{a^{\dagger}\},\{a\})$ is
	\be
	{H}(\{a^{\dagger}_i,a_i\})=-\frac{1}{2}\sum_i\sum_jK(i,j) (a^{\dagger}_{i+j}-a^{\dagger}_i a ^{\dagger}_j)a_ia_j.
	\ee 
	The solution of Eq.~(\ref{sch}) is
	\begin{equation}
	\ket{\psi(t)}=e^{-{H}t}\ket{M,0,\ldots,0},
	\end{equation}
	where, initially ($t=0$), there are $M$ particles of mass $1$, \textit{i.e.}, $\ket{\psi(0)}=\sum_{\vec{N}}\delta_{\vec{N},\{M,0,0\ldots 0\}}\ket{\vec{N}}$.
	
	Multiplying both sides of the equation on the left by an arbitrary state $\bra{\vec{L}}$ and using the relation $\braket{\vec{L}|\vec{N}}=\frac{1}{\vec{N}!}\delta_{\vec{L},\vec{N}}$,
	\begin{equation}
	P(\vec{N},t)=\frac{\braket{\vec{N}|\psi(t)}}{\vec{N}!}.\label{supP}
	\end{equation}

	By definition, the probability that $N$ clusters survive at time $t$, $P(M,N,t)$, can be expressed in terms of $P(\vec{N},t)$:
	\begin{equation}
	P(M,N,t)=\sum_{\vec{N}}P(\vec{N},t)\delta\left(\sum_{i=1}^{k^*}N_i-N\right),
	\end{equation} where $k^*=M-N+1$ is the maximum mass that can be formed after time $t$, given that $N$ clusters remain. 
	Substituting the expression for $P(\vec{N},t)$ from Eq.~(\ref{supP}), and multiplying and dividing by $N!$, we obtain,
\begin{equation}
P(M,N,t)=\frac{1}{N!}\braket{\vec{0}|\sum_{k_1} a_{k_1}\sum_{k_2} a_{k_2}\dots \sum_{k_N}a_{k_N} e^{-{H}(a^{\dagger},a)t} |\psi(0)}.
\end{equation}
We ensure mass conservation by introducing a constrained sum:
\begin{equation}
P(M,N,t)=\frac{1}{N!}\braket{\vec{0}|\sideset{}{'}\sum_{k_i=1}^{k^*}\prod_{i=1}^{N} a_{k_i} e^{-{H}(a^{\dagger},a)t} |\psi(0)}.\label{braket}
\end{equation}
where $\prime$ on the summation denotes the constrained sum.

	In order to write Eq.~(\ref{braket}) as a path integral, the evolution operator $e^{-\widehat{H}(a^{\dagger},a)t}$ is split into a product of the evolution operators $e^{-H\epsilon}$ for infinitesimal times $\epsilon$, in the limit $\epsilon\to 0$:
	\begin{equation}
	P(M,N,t)=\frac{1}{N!}\lim_{\epsilon\to 0}\braket{\vec{0}|\sideset{}{'}\sum_{k_i=1}^{k^*}\prod_{i=1}^{N} a_{k_i}\prod_{n=1}^{t/\epsilon} e^{-{H}(a^{\dagger},a)\epsilon} |\psi(0)}\label{seq:P}
	\end{equation}	
Then, we insert identity operators $\textbf{I}$ for every infinitesimal evolution $e^{-H\epsilon}$ in terms of coherent states, $\ket{z}$ and their complex conjugates. The coherent state $\ket{z}$ and $\textbf{I}$ are defined as follows,
	\begin{eqnarray}
	&&a_i\ket{z}=z_i\ket{z},\label{seq:z}\\
	&&\bra{z}a_i^{\dagger}=\bra{z}\tilde{z}_i,\label{seq:tz}\\
	&&\textbf{I}=\int\prod_{i} \frac{dz_id\tilde{z}_i}{\pi}e^{-\sum_i z_i\tilde{z}_i}\ket{z}\bra{z}\label{seq:I},
	\end{eqnarray}	 
	where $\tilde{z}_i$ is the complex conjugate of $z_i$ ($|z_i|^2=z_i\tilde{z}_i$), and $\ket{z}$ is written in terms of creation operators as
	\begin{equation}
	\ket{z}=e^{-\frac{1}{2}\sum_m|z_m|^2}e^{\sum_m z_m a_m^{\dagger}}\ket{0}.
	\end{equation}

	Using Eqs.~(\ref{seq:z}), (\ref{seq:tz}), and (\ref{seq:I}) in Eq.~(\ref{seq:P}), $P(M,N,t)$ is written as follows,
	\begin{align}
	&P(M,N,t)= \frac{1}{N!}\sideset{}{'}\sum_{k_i=1}^{k^*}\int \{D\tilde{z}_i(t)\}\{Dz_i(t)\} \exp\left(-{\int_{0}^{t_f}\!\!\!\!dt\left[\sum_i\tilde{z}_i\dot{z}_i+H\right]+M\ln \tilde{z}_1(0)+M\ln\prod_{n=1}^Nz_{k_n}(t_f)}\right), \label{eq:p}
	\end{align}
	Notice that the exponent is invariant under the transformation $z_m\to c^m z_m$, $\tilde{z}_m\to c^{-m}\tilde{z}_m$, where $c$ is a constant. Hence, we can choose $c=\tilde{z}_1(0)$, in order to eliminate the term $M\ln \tilde{z}_1(0)$. The choice of $c$ further implies that $z_1(0)=M$. 
	
	\subsection{Scaling to obtain the action}
	Let $\tau = M^{\alpha} t$ and $z_i(\tau)\to M^{\beta}z_i(t)$. We observe that there is no scaling possible for $\tilde{z}_i$, as it is a dimensionless quantity [see Eq.~(\ref{eq:energy}) of main text]. Scaling the integrand in the exponential in Eq.~(\ref{eq:p}), we find that $\alpha=1, \beta=-1$ and $H\to H/M^2$. Using the Stirling formula to write 
\begin{equation}
ln N!=N\ln N-N,
\end{equation} 
and using the Laplace method in the limit $M\to\infty$,  Eq.~(\ref{eq:p}) is 
	\begin{align}
	&P(M,\phi,\tau_f)\approx \sideset{}{'}\sum_{k_i=1}^{k^*} e^{-MS(\phi,\tau_f;\{z_i,\tilde{z}_i\})}, \label{eq:p1}
	\end{align}
where $\phi=N/M$, $\tau=Mt$, and the action can be written as
\begin{equation}
S(\phi,\tau_f;\{z_i,\tilde{z}_i\})={\int_{0}^{\tau_f}d\tau\left[\sum_{i}\tilde{z}_i\dot{z}_i+H\right]+1-\frac{\ln\prod_{n=1}^Nz_{k_n}(\tau_f)}{M}+\phi\ln\phi-\phi}.
\end{equation}	

In order to explicitly obtain the rate function, we calculate $z_{m}(\tau)$ for the constant, sum and product kernels in the following sections.
	\subsection{Energy is a constant of motion}
	Using the Euler-Langrange equations derived by minimizing the action [see Eqs.~(\ref{z}) and (\ref{tz}) of main text], we prove that $H$ is a constant of motion. For an arbitrary $K(i,j)$,
\begin{equation}
\frac{dH}{dt}=\frac{\partial H}{\partial z}\dot{z}+\frac{\partial H}{\partial \tilde{z}}\dot{\tilde{z}}.\label{eq:e0}
\end{equation} 
Substituting the Euler-Lagrange equations in Eq.~(\ref{eq:e0}), we prove that $\frac{dH}{dt}=0$.

\section{Constant Kernel}
The instanton trajectory for a given energy $E$, for the constant kernel, can be obtained by solving the equation
\begin{equation}
\frac{dn}{d\tau}=E-\frac{n^2}{2},\label{seq:n}
\end{equation}
where $E<0, E=0,E>0$ correspond to $\tau_f<\tau_{typ},\tau_f=\tau_{typ}, \tau_f>\tau_{typ}$ respectively for a given $\phi$. 

When $E<0$, let $E=-p^2/2$. Then,
\begin{equation}
\int_{n(0)=1}^{n(\tau)}\frac{dn}{n^2+p^2}=-\frac{1}{2}\int_{0}^{\tau}d\tau,
\end{equation}
which can be solved to give
\begin{equation}
n(\tau)=-p\tan\frac{p(\tau-\tau_0)}{2},\label{seq:E<0}
\end{equation}	
where $\tau_0$ and $p$ can be fixed using the initial and final conditions.
For final time $\tau_f\to 0$, $E$ is large and negative. Hence, 
\begin{equation}
E\approx-\frac{1-\phi}{\tau_f}.
\end{equation}

When $E=0$, Eq.~(\ref{seq:n}) is easily solved to give
\begin{equation}
n(\tau)=\frac{1}{1+\tau/2}.
\end{equation}
This solution corresponds to the typical trajectory.

When $E>0$, let $E=p^2/2$. Since $n(\tau)$ is a decreasing function of time, $E<n^2/2$. Following the procedure described to obtain Eq.~(\ref{seq:E<0}), 
\begin{equation}
n(\tau)=p\coth \frac{p(\tau-\tau_1)}{2}.
\end{equation}

The equation for $Y(x,\tau)$ is similar to the equation for $n(\tau)$, and can be solved following the above procedure, for all $E$. 

\section{Sum Kernel}
In order to calculate $z_m(\tau)$ for the sum kernel, we rewrite $z_m(\tau)=c_m(\tau)\exp\big(-\int d\tau \frac{mn(\tau)+1}{2}\big)$. Equation~(\ref{z}) of the main text now becomes
\begin{equation}
\frac{dc_m}{d\tau^{\prime}}=\frac{1}{2}\sum_j mc_jc_{m-j},
\end{equation} 
where $d\tau^{\prime}/d\tau=\exp(-\tau/2)/2$. Further, using the ansatz $c_m(\tau^{\prime})=a_m \tau^{\prime m-1}$, where $a_m$ is a function of mass alone, and solving the resulting equation, we obtain 
\begin{equation}
c_m(\tau)=\frac{m(1-e^{-\tau/2})^{m-1}z_1(0)^m}{m!},
\end{equation}
and hence,
\begin{equation}
z_m(\tau)=\frac{m^{m-1}a_1^m}{m!}(1-e^{-\tau/2})^{m-1}e^{-\int d\tau^{\prime} \frac{mn+1}{2}},\label{seq:zm}
\end{equation}
where $a_1=1$ from the initial condition.

The combinatorial factor for the sum kernel (Eq.~(\ref{eq:sumcomb}) of main text) is calculated as follows. We write the constrained sum as the integral form of a $\delta$-function,
\begin{equation}
\delta\bigg(\sum_i k_i-M\bigg)=\oint \frac{dx}{2\pi ix} x^{-M+\sum_{i=1}^N k_i}.
\end{equation}
Then,
\begin{equation}
\!\!\!\sideset{}{'}\sum_{k_i=1}^{k^*} \prod_{n=1}^{N}\frac{k_n^{k_n-1}}{k_n!}\!\!=\oint \frac{dx}{2\pi ix} x^{-M} \bigg(h(x)\bigg)^N,\label{eq:W}
\end{equation}
where $h(x)=\sum_{k=1}^M \frac{k^{k-1}x^k}{k!}$. Further, $h(x)$ can be written in terms of the Lambert $W-$function,
\begin{equation}
h(x)=-W(-x).
\end{equation} 
Making a change of variable from $x$ to $W(-x)$ in Eq.~(\ref{eq:W}),
\begin{equation}
\sideset{}{'}\sum_{k_i=1}^{k^*} \prod_{n=1}^{N}\frac{k_n^{k_n-1}}{k_n!}=\oint \frac{dW}{2\pi iW} (-1)^{N-M} W^{-M+N}(W+1)e^{-MW}.
\end{equation}
Writing the integrand as $e^{-Mf(W)}$ and using the Laplace method to find the minimum of $f(W)$, we obtain the combinatorial factor in terms of $\phi$ and $M$.
The large deviation function, Eq.~(\ref{eq:sum}) of the main text, is derived by substituting the combinatorial factor, and Eq.~(\ref{seq:zm}), and is found to be in excellent agreement with Monte Carlo simulations, as shown in Fig.~\ref{fig:02} of the main text.

\section{Product kernel}
\begin{figure}
	\includegraphics[width=\linewidth]{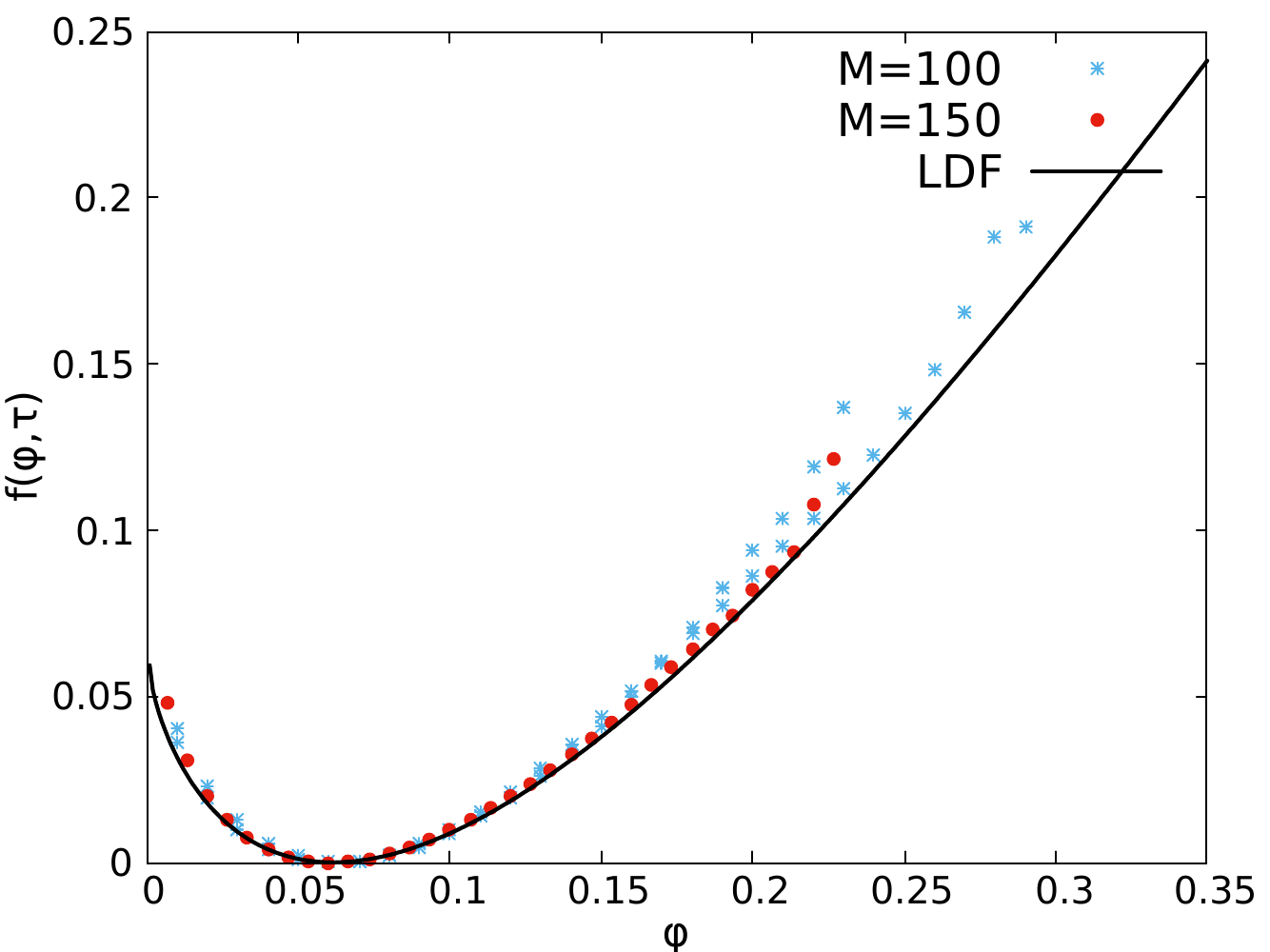}
	\caption{$f(\phi,\tau)$ with respect to $\phi$ for $\tau=3$ for $M=100, 150$.}
	\label{fig:tau3}
\end{figure}
Product kernel is an exactly solvable kernel which exhibits gelation, a phenomenon in which a single macroscopic cluster with size of the order of total mass $M$ forms at $\tau>1$, and 'eats up' the smaller clusters. The Smoluchowski equation fails when the gel appears, and the total mass is no longer conserved, necessitating the use of approximate techniques to obtain the mass distribution\cite{krapivsky2010kinetic,leyvraz2003scaling}. The exact typical mass distribution for the product kernel was obtained by Lushnikov~\cite{lushnikov2006gelation,lushnikov1973evolution,lushnikov1978coagulation}.

We now compute the LDF for the product kernel. If we proceed in the same way as we did for the constant and sum kernels, we obtain an LDF which works only within the regime where the Smoluchowski equation is valid. In order to obtain the correct LDF, we start by rewriting $H$ as given in Eq.~(\ref{eq:H'}) of the main text. Upon introducing coherent states and solving as before, we obtain 
\begin{equation}
P(M,\phi,\tau_f)\approx \lim_{M\to\infty}\sideset{}{'}\sum_{k_1,k_2\dots k_N=1}^{k^*} \exp\left[-M\left(\phi\ln\phi+1-\phi-\frac{\sum_{n=1}^N\ln z_{k_n}}{M}+ 
\!\!\int_{0}^{\tau_f}\!\!d\tau \Big[\sum_m \tilde{z}_m\dot{z}_m+E\Big]\right)\right]\!\!,
\end{equation}
where 
\begin{equation}
H=-\frac{1}{2}\sum_{i=1}^{k^*} \sum_{j=1}^{k^*} ij\tilde{z}_{i+j}z_iz_j+ \frac{1}{2}-\sum_{j=1}^{k^*}\frac{j^2\tilde{z}_jz_j}{2M^2}.
\end{equation}
In order to compute $z_m(\tau)$, we solve the unscaled Euler-Lagrange equation for $z_m(t)$ (see Eq.~(\ref{eq:pz}) of main text). Let 
\begin{equation}
G(x,t)=\sum_m z_m(t)x^m.
\end{equation}
Then,
\begin{equation}
\frac{\partial G}{\partial t}=\frac{1}{2}\left(x\frac{\partial G}{\partial x}\right)^2-Mx\frac{\partial G}{\partial x}+\frac{1}{2}x\frac{\partial}{\partial x}\left(x\frac{\partial G}{\partial x}\right),~~~G(x,0)=z_1(0)x.
\end{equation}
Making the Cole-Hopf transformation $\ln D(p(x,t),t)=G(x,t)$, where $p(x,t)=xe^{-Mt}$, we obtain
\begin{equation}
\frac{\partial D}{\partial t}=\frac{1}{2}\frac{\partial}{\partial p}\left(p\frac{\partial D}{\partial p}\right),\label{eq:D}
\end{equation}
with the initial condition $D(x,0)=e^{z_1(0)x}$.
Let \begin{equation}
D(p,t)=\sum_m a_m(t) f(m)p^m. 
\end{equation}
Then, substituting in Eq.~(\ref{eq:D}) and matching the coefficients of $p^m$ on both sides of the equation,
\begin{equation}
\frac{da_m}{dt}=\frac{m^2 a_m}{2},
\end{equation}
which can be solved to obtain 
\begin{equation}
a_m(t)=ce^{\frac{m^2 t}{2}}.\label{eq:a}
\end{equation}
We use the initial condition to obtain $f(m)$,
\begin{equation}
f(m)=\frac{z_1(0)^m}{m!}.\label{eq:f}
\end{equation}
Combining Eqs.~(\ref{eq:a}) and (\ref{eq:f}), 
\begin{equation}
D(p,t)=\sum_m \frac{e^{\frac{m^2 t}{2}}(z_1(0)p)^m}{m!}.
\end{equation}
In order to extract $G(x,t)$ from $D(p,t)$, we use Knuth identity~\cite{knuth1998linear}: 
\begin{equation}
\ln\sum_{m=1}^{\infty}\frac{x^{m(m-1)/2}z^m}{m!}=\sum_{m=1}^{\infty}\!\!\frac{(x-1)^{m-1}F_{m-1}(x)z^m}{m!},
\end{equation}
where $F_{m-1}(x)$ are known as Mallows-Riordan polynomials, and obey the following recursion relation:
\begin{equation}
F_{m}(x)=\sum_{l=1}^m {m-1\choose{l-1}}\sum_{i=0}^{l-1}x^iF_{l-1}(x)F_{m-l}(x).
\end{equation}
Converting $D(p,t)$ in terms of $x$, and equating the coefficients of $x^m$ on both sides of the equation, we  obtain\begin{equation}
z_m(t)=\frac{(e^{t}-1)^{m-1}F_{m-1}(e^{t})M^{m}e^{m(-Mt+t/2)}}{m!},
\end{equation}

Scaling $z_{m}(t)\to z_m(\tau)M$ and $t\to\tau/M$, we obtain
\begin{equation}
z_m(\tau)=\frac{(e^{\tau/M}-1)^{m-1}F_{m-1}(e^{\tau/M})M^{m-1}e^{m(-\tau+\tau/2M)}}{m!}.
\end{equation} 
Substituting the expression for $z_m(\tau)$ in Eq.~(\ref{eq:actionfunctional}) of the main text, we obtain the final LDF for the product kernel. Comparison with Monte Carlo results shows that the LDF is in good agreement with the simulations. Unlike the other simulations, Fig.~(\ref{fig:03})(b) for the LDF with respect to $\phi$ for $\tau=1.4$ contains simulations for small $M$ values. This is due to the limitation that the simulations are computationally expensive and consume a large amount of time. But it can be seen that even for such cases, the agreement with the analytical LDF is fairly good, and improves for larger $M$. Additionally, the LDF shows good agreement with simulation results for much larger values of $\tau$ as well, as shown in Fig.~\ref{fig:tau3}.

\section{Numerical algorithm}
We briefly describe the Monte Carlo algorithm designed to numerically determine $P(M,N,t)$ for any given collision kernel. A trajectory that contributes to $P(M,N,t)$ consists of $C=M-N$ collisions, and is defined by the waiting times between collisions, sequence of collisions and the number of collisions. No collision occurs after the final waiting time. At $t=0$, there are $M$ particles of mass $1$. A trajectory is evolved by reassigning a waiting time with probability $p_1$, reassigning a collision with probability $p_2$, or modifying the number of collisions with probability $p_3$, such that $p_1+p_2+p_3=1$. Each of these moves is described below. 

\paragraph{\textbf{Reassign waiting times}}: Keeping the number of collisions $C$, total time for $C$ collisions, and sequence of collisions fixed, a collision $i$ is picked at random. $\Delta t_i$ is the waiting time between the $i^{th}$ and $(i+1)$-th  collisions, and $\lambda_i$ is the total rate of collision of the $i^{th}$ configuration. The waiting times $\Delta t_i$ and $\Delta t_{i+1}$ are modified, keeping the sum $\Delta t_i+\Delta t_{i+1}$ constant, such that the total time remains unchanged~\cite{dandekar2023monte}.
 
\paragraph{\textbf{Reassign a collision}}: Keeping the number of collisions and the total time for $C$ collisions fixed, a collision $i$ is chosen at random, and reassign according to the rules listed in~\cite{dandekar2023monte}.

\paragraph{\textbf{Add/delete collision}}: With equal probability, a collision is added or deleted.

In order to add a collision, two masses  $m_1$ and $m_2$ are selected at random from the configuration resulting from the final collision. The collision rate of these masses are calculated, and used to generate the waiting time for the $(C+1)$-th collision, $\Delta t_{C+1}$. 

In order to delete a collision, the final configuration is made equal to the previous one, and the waiting times are modified such that $\Delta t_{C}=\Delta t_{C}+\Delta t_{C+1}$. 

Addition and deletion of a collision are performed in such a way that the principle of detailed balance is satisfied. Suppose the old trajectory consists of $C$ collisions, and the new trajectory consists of $C+1$ collisions. The probability of the old and new trajectories are $P(C)$ and $P(C+1)$ respectively, and the weights associated with adding a collision, and deleting a collision, are $Wt(C\to C+1)$ and $Wt(C+1\to C)$ respectively.  The condition for detailed balance is:
\begin{equation}
P(C)Wt(C\to C+1)=P(C+1)Wt(C+1\to C).
\label{db}
\end{equation}
The old trajectory has the probability
\begin{equation}
P(C)=\frac{1}{\lambda_C}\left[\prod_{i=0}^{C}\lambda_i e^{-\lambda_i\Delta t_i}\prod_{i=1}^{C}P(\xi_i)\right] e^{WC},\label{seq:C}
\end{equation} 
where $\xi_i$ denote the possible configurations and $W$ is a bias parameter.
The new trajectory has the probability
\begin{equation}
P(C+1)=\frac{1}{\lambda_{C+1}}\left[\prod_{i=0}^{C+1}\lambda_i e^{-\lambda_i\Delta t_i}\prod_{i=1}^{C+1}P(\xi_i)\right] e^{W(C+1)}.\label{seq:C+1}
\end{equation} 
The protocol for adding a collision after the $C-$th collision is the following:
\begin{itemize}
\item Let there be L possible mass pairs which can collide, \textit{i.e.,} $j=1,2,\dots L$. The $j-$~th mass pair is chosen with probability 
\begin{equation}
\frac{K(m_1,m_2)n_{m_1}n_{m_2}}{\sum_{i,j}K(m_i,m_j)n_{m_i}n_{m_j}}.
\end{equation}
\item Choosing the mass pair fixes the collision rate as $\lambda^{(j)}_{C+1}$. Using this rate, we choose $\Delta t^{\prime j}_C$ such that $\Delta t^{\prime j}_C+\Delta t^{\prime}_{C+1}=\Delta t_C$, from the distribution 
\begin{equation}
P(\Delta t^{\prime j}_C)=\frac{(\lambda^{\prime j}_{C+1}-\lambda_C)e^{(\lambda^{\prime j}_{C+1}-\lambda_C)\Delta t^{\prime j}_C}}{e^{(\lambda^{\prime j}_{C+1}-\lambda_C)\Delta t_C}-1}.
\end{equation} 
\item Now, the weight of adding a collision is 
\begin{equation}
Wt(C\to C+1)=P(\xi_j)P(\Delta t^{\prime j}_C)\label{seq:WC}
\end{equation}.
\end{itemize}

The protocol for deleting a collision just involves deleting the final configuration and setting $\Delta t_C=\Delta t^{\prime}_C+\Delta t^{\prime}_{C+1}$. That is, 
\begin{equation}
Wt(C+1\to C)=1/\kappa,\label{seq:WC+1}
\end{equation} where $\kappa$ is a constant.

Substituting Eqs.~(\ref{seq:C}), (\ref{seq:C+1}), (\ref{seq:WC}) and (\ref{seq:WC+1}) into Eq.~(\ref{db}), we obtain 
\begin{equation}
\kappa=\frac{\lambda_C e^W (1-e^{-(\lambda_{C+1}-\lambda_C)\Delta t_C})}{\lambda_{C+1}-\lambda_C}.
\end{equation}

That is, addition of a collision is accepted with probability $\min(\kappa,1)$, and deletion of a collision with probability $\min(1/\kappa,1)$.

Implementing the algorithm described, we see that the numerical results are in excellent agreement with the analytical LDF, as seen in Fig~\ref{fig:tau3}.